\documentstyle[iopfts]{ioplppt}
\begin{document}
\title{Ringing the eigenmodes from compact manifolds}

\author{Neil J. Cornish and Neil G. Turok}

\address{Department of Applied Mathematics and Theoretical
Physics, University of Cambridge, Silver Street, Cambridge CB3 9EW, UK}

\begin{abstract}
We present a method for finding the eigenmodes of the Laplace operator
acting on any compact manifold. The procedure can be used to
simulate cosmic microwave background fluctuations in multi-connected
cosmological models. Other applications include studies of chaotic
mixing and quantum chaos.
\end{abstract}

\section{Introduction}

Much of physics boils down to solving differential equations subject
to certain boundary conditions. Waves in a box are a classic example.
The box we have in mind here is some closed manifold, and the waves
are those of a massless scalar field. This picture
arises naturally when studying density perturbations in a
multi-connected universe\cite{ll}. The mathematical problem can be
stated: Find all square
integrable functions $\Psi_q({\bf x})$ that satisfy the partial
differential equation
\begin{equation}
(\nabla + q^2 )\Psi_q({\bf x}) = 0 \, .
\end{equation}
Here $\nabla$ is the Laplace operator on some closed manifold $\Sigma$
and the constant $q$ is an eigenvalue of the Laplacian. The
complexity of the problem is controlled by the geometry of $\Sigma$.
When $\Sigma$ is $n$-dimensional Euclidean space, $E^n$, modulo some
discrete group of covering transformations, $\Gamma$, it is a simple
matter to write down analytic expressions for the eigenmodes
$\Psi_q({\bf x})$. In contrast, if $\Sigma$ describes some compact
hyperbolic manifold, $H^n / \Gamma$, then the eigenmodes cannot be
expressed in closed analytic form. This difficulty is
closely related to the chaotic behaviour of geodesic
flows on compact negatively curved spaces\cite{berry}. Here we
describe a numerical solution to the problem based on Fourier filtering
solutions to the scalar wave equation $\Box \Psi =0$. While our
approach works for any topology, we will focus on hyperbolic manifolds
as these are of the most interest to cosmology.

\section{Ringing out the modes}

Rather than attack Laplace's equation directly, we begin by
introducing a ficticious time dimension $t$, thereby lifting the
problem to solving the wave equation
\begin{equation}\label{wave}
\left( { \partial^2  \over \partial t^2} - \nabla \right)
\Psi_q(t,{\bf x}) = 0 \, ,
\end{equation}
on the Lorentzian manifold ${\cal M}=R \times \Sigma$. Consistent with
this picture we adopt a product ansatz for the eigenmodes:
\begin{equation}
\Psi_q(t,{\bf x}) = \exp(-i\omega_q t)\Psi_q({\bf x}) \, .
\end{equation}
The eigenfrequency $\omega_q$ is fixed by (\ref{wave}) to equal $q$.
Assuming for now that the spatial boundary
conditions have been properly enforced, we can evolve the initial data
$\Psi(0,{\bf x})$, $\partial_t \Psi(t,{\bf x})\vert_0 = 0$ according
to (\ref{wave}) to find $\Psi(t,{\bf x})$. By Fourier transforming in
time:
\begin{equation}
a_{\omega}({\bf x}) = \int_{0}^{\infty} \Psi(t,{\bf x}) e^{i \omega t}
dt \, ,
\end{equation}
and calculating the power spectrum
\begin{equation}
P(\omega) = \int_{\Sigma} \vert a_{\omega} \vert \, \sqrt{g}\, 
{d^nx} \, ,
\end{equation}
we are able to isolate the eigenfrequencies $\omega_q$. These are
located at local maxima of $P(\omega)$. Once the eigenfrequencies
are known the individual spatial eigenmodes can be extracted:
\begin{equation}
\Psi_q({\bf x}) = \lim_{T \rightarrow \infty}{1 \over T}
\int_{0}^{T} \Psi(t,{\bf x}) e^{i q t} dt \, .
\end{equation}
In practice the integration time $T$ will be finite, so we can only
resolve modes separated in frequency by at least $\Delta \omega = 2\pi /T$.

The main difficulty in implementing the above procedure
stems from the complicated periodic boundary conditions that 
are imposed by the topology. The remainder of this
paper is devoted to describing a numerical solution to this problem,
and illustrating how it works by finding the eigenmodes for a genus 2
surface with constant negative curvature.

\section{The double doughnut}

The manifold we will focus on is the two-hole doughnut $\Sigma=
H^2/\Gamma$, where
$\Gamma$ is a discrete subgroup of $SO(2,1)$ with presentation
\begin{equation}
\Gamma=\{ g_0,\, g_1,\, g_2,\, g_3\, : g_0 g_1^{-1}g_2 g_3^{-1}
g_0^{-1} g_1 g_2^{-1} g_3 \} \, .
\end{equation}
The group generators have the $SO(2,1)$ matrix representation\cite{balazs}
\begin{equation}
g_j = R_j
\left( \begin{array}{ccc}
\cosh\eta & \sinh\eta & 0 \\
\sinh\eta & \cosh\eta & 0 \\
0 & 0 & 1
\end{array} \right) R_j^{-1}
 \, ,
\end{equation}
where $\eta = 2\, {\rm arccosh}(1+\sqrt{2})$ and
\begin{equation}
R_j=\left( \begin{array}{ccc}
1 & 0 & 0 \\
0 & \cos(j\pi/4) & -\sin(j\pi/4) \\
0 & \sin(j\pi/4) & \cos(j\pi/4)
\end{array} \right) \, .
\end{equation}
The double doughnut can be obtained by gluing together opposite faces
of a regular octagon with dihedral angles $\pi/4$. The octagonal
fundamental domain (FD) is drawn in Fig.~1 using the Poincar\'{e} disc
model for $H^2$.

\newpage

\
\begin{figure}[h]
\vspace*{60mm}
\includegraphics{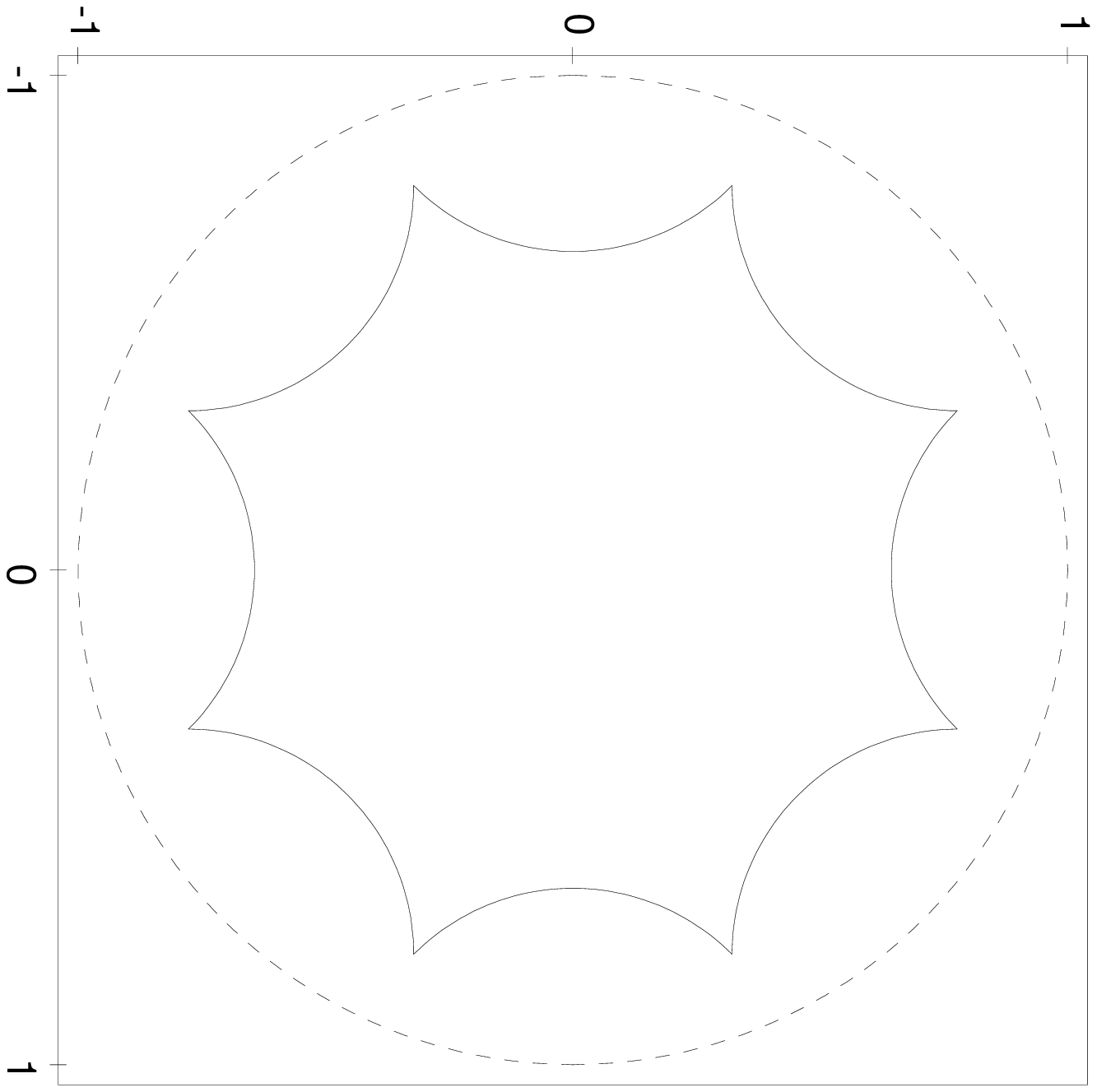}
\vspace*{5mm}
\caption{The fundamental domain for the double doughnut (shown bordered by a
solid line) in the Poincar\'{e} disk (bordered by a dashed line).}
\end{figure}

\vspace*{-5mm}
\begin{picture}(0,0)
\put(193,35){$x$}
\put(90,135){$y$}
\end{picture}

The double doughnut has area $4\pi$ and diameter $d=2\, {\rm arctanh}
(2^{-1/4})\approx 2.448$. All distances are quoted in units of the
curvature radius. The largest simply connected circle that can be
drawn inside $\Sigma$ has radius $\eta_-= {\rm arccosh}(1+\sqrt{2})$,
and the smallest circle to fully enclose the fundamental domain has
radius $\eta_+ = 2 {\rm arccosh}(\sqrt{2+\sqrt{2}})$. The light
crossing time varies in the interval $2 \eta_- \leq t_c \leq 2\eta_+$, and has
the mean value $t_\lambda = \sqrt{4\pi} \approx 3.545$. This
is the Lyapunov time for chaotic flows on $\Sigma$, and
represents the characteristic dynamical timescale for our system.

\section{Solving the wave equation}

The dynamics is described by a massless scalar field $\phi$ with action
\begin{equation}
S = \int \sqrt{-g} g^{\mu\nu} \partial_\mu \phi \partial_\nu \phi\,
d^3 x \, ,
\end{equation}
evolving on the Lorentzian manifold $(R\times \Sigma, g)$. Choosing a
Poincar\'{e} metric for $H^2$ we have
\begin{equation}
ds^2 = -dt^2 + { 4 ( dx^2 +dy^2 ) \over (1-x^2-y^2)^2} \, .
\end{equation}
The coordinate distance $r=(x^2+y^2)^{1/2}$ is related to the
proper distance $\eta$ by $r={\rm tanh}(\eta /2)$. 
To evolve the system numerically we begin by discretising time and
space such that $x = i\, \Delta x$, $y=j\, \Delta y$ and $t = k\, \Delta t$.
The discretised action then reads
\begin{eqnarray}
\hspace*{-0.4in} S_\Delta
= && \hspace*{-0.1in} \sum_{i,j,k}
\left[ { (\phi(i+1,j,k)-\phi(i,j,k) )^2 \over \Delta x^2}
+{ (\phi(i,j+1,k)-\phi(i,j,k) )^2 \over \Delta y^2} \right. \nonumber
\\
&&  \left.
- { 4 \over (1-(i\Delta x)^2-(j \Delta y)^2)^2}
{ (\phi(i,j,k+1)-\phi(i,j,k) )^2 \over \Delta t^2}\right]\Delta x
\Delta y \Delta t \, .
\end{eqnarray}
The equations of motion that follow by varying $S_\Delta$ with respect to
$\phi(i,j,k)$ are:
\begin{equation}\label{eveq}
4\delta_k^2 \phi(i,j,k) =  (1-(i\Delta x)^2-(j \Delta y)^2)^2 
\left[ \delta_i^2 \phi(i,j,k) + \delta_j^2 \phi(i,j,k) \right] \, 
\end{equation}
where
\begin{equation}
\delta_i^2 f(i) = { f(i+1) - 2 f(i) + f(i-1) \over (\Delta x^i)^2 } \, .
\end{equation}

\
\begin{figure}[h]
\vspace*{65mm}
\includegraphics{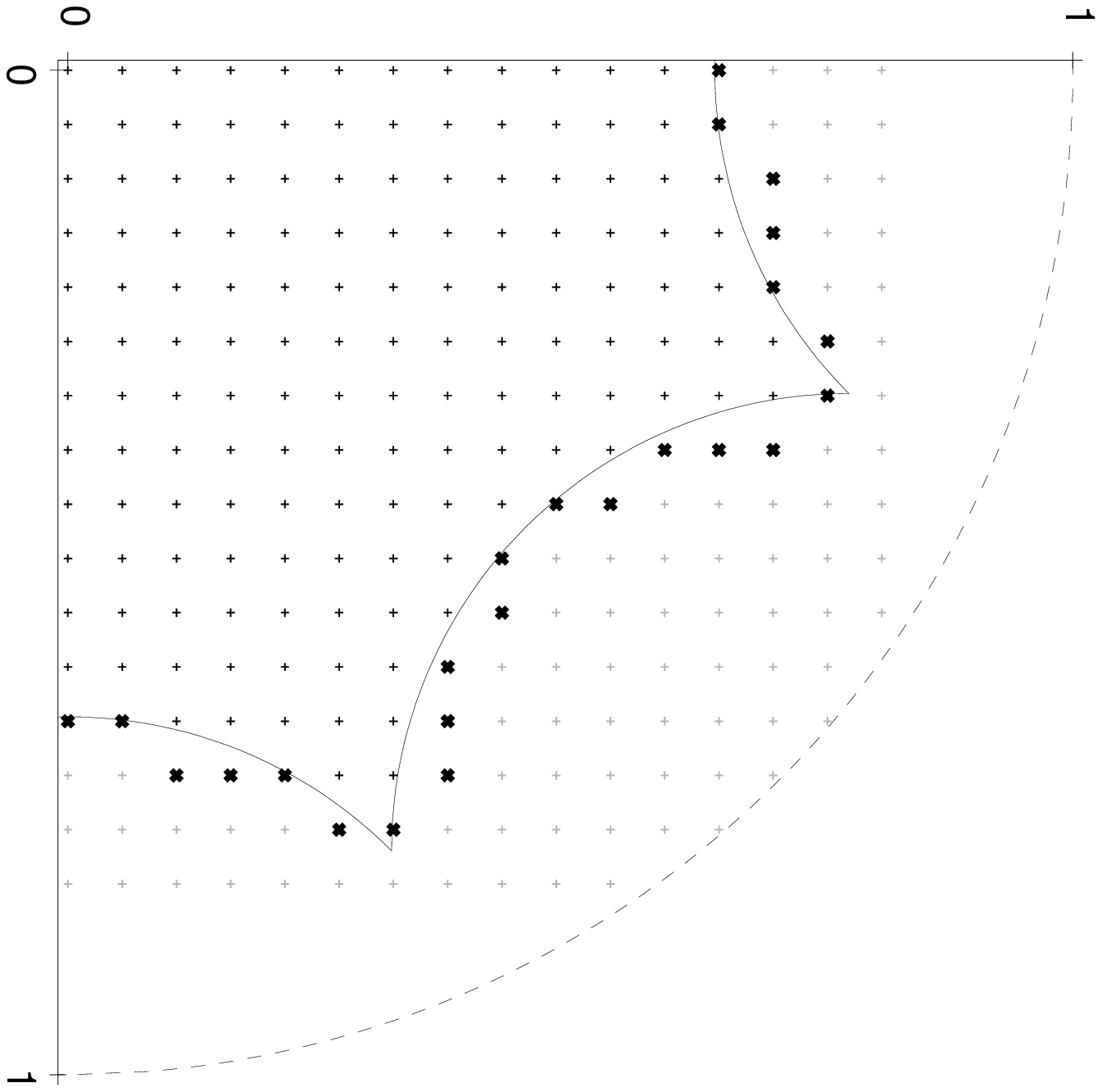}
\includegraphics{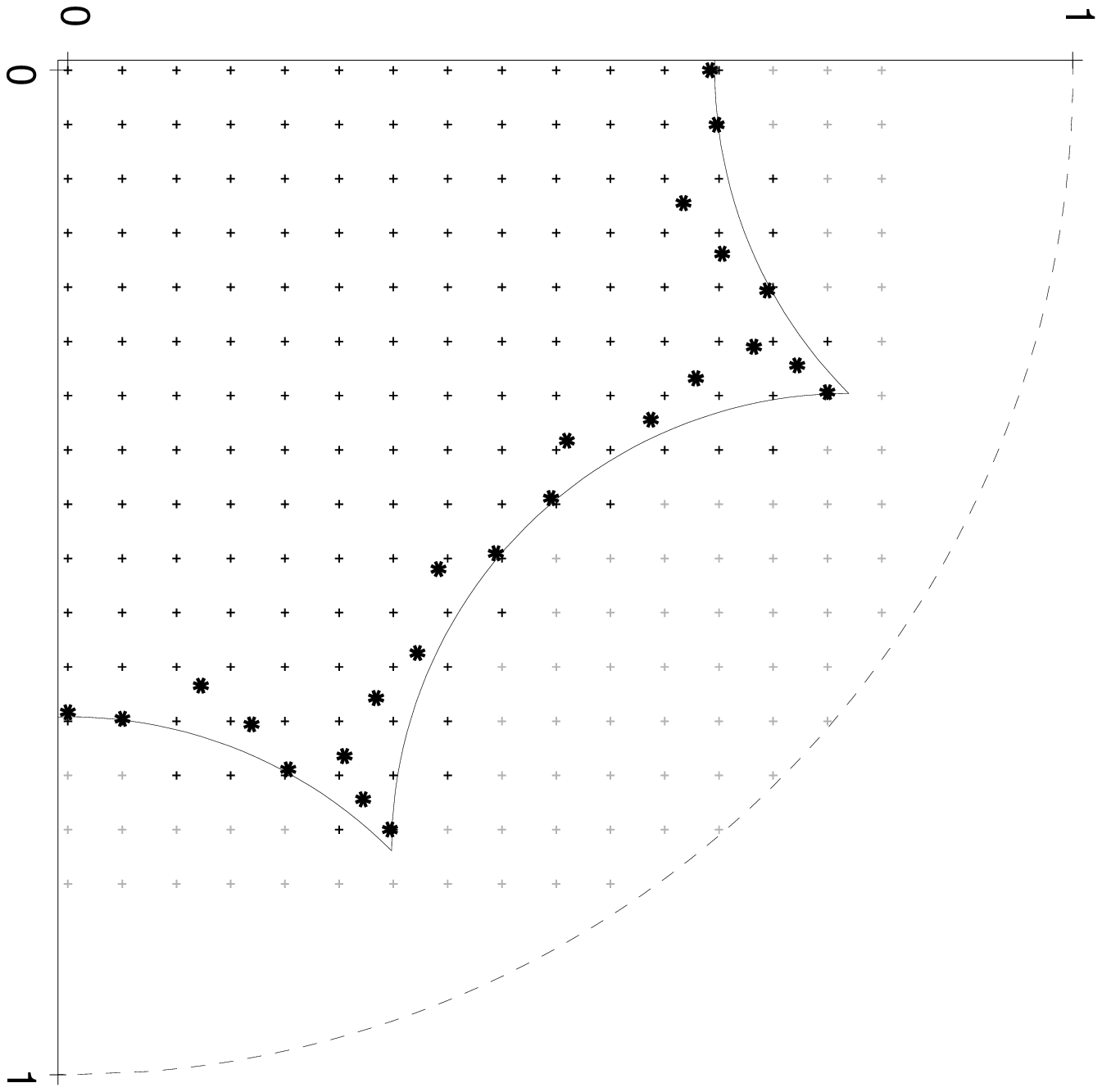}
\vspace*{5mm}
\caption{(a) The grid points are first sorted into masters (dark
crosses), slaves (heavy dots) and freemen (light crosses). (b) The
slave images (heavy dots) are located inside the FD.}
\end{figure}

\vspace*{-2mm}
\begin{picture}(0,0)
\put(70,57){(a)}
\put(260,57){(b)}
\end{picture}

Our next task is to enforce periodic boundary conditions at the edge
of the fundamental domain. To do this we begin by sorting all points
in the spatial grid into those inside and outside the fundamental
domain. There is a simple sorting algorithm that works for all
topologies. In what follows, $\vert {\bf x} \vert$ denotes the proper
distance from the origin.

\centerline\underline{Algorithm 4.0. Point Sorting}

\begin{enumerate}
\item If $\vert {\bf x} \vert < \eta_-$ then the point lies inside the
FD. If $\vert {\bf x} \vert > \eta_+$ then the point lies outside the FD.
\item For $\eta_- < \vert {\bf x} \vert <\eta_+$, act on
${\bf x}$ by all $n$ group generators and their inverses
to form the $2n$ image points ${\bf x}'_\pm i= g_i^{\pm 1} {\bf x}$.
\item If $\vert {\bf x}'_{\pm i} \vert < \vert {\bf x} \vert $ for any
$i$, then ${\bf x}$ lies outside the fundamental domain, else ${\bf x}$
is inside the fundamental domain.
\end{enumerate}

Only the ``master'' points inside the FD need to be evolved.
However, forming the second derivatives $\delta_i^2 \phi$ and
$\delta_j^2 \phi$ for the masters often requires knowledge of field
values outside the FD. We refer to all points outside the FD that are
within one grid point of a master as ``slaves''. Points that are
neither masters nor slaves play no part in the numerical evolution and
are designated ``freemen''. Each
slave has a unique image inside the FD. We can find the
position of these images using the following algorithm.

\centerline\underline{Algorithm 4.1. Locating the fundamental image}

\begin{enumerate}
\item Act on ${\bf x}$ by all $n$ group generators and their inverses
to form the $2n$ image points ${\bf x}'_\pm i= g_i^{\pm 1} {\bf x}$.
\item Find the image point ${\bf x}'_\pm i$ nearest the
origin and call it ${\bf x}'$.
\item If $\vert {\bf x}' \vert < \vert {\bf x} \vert $
then let ${\bf x} = {\bf x}'$ and go to (i), else ${\bf x}$ is the
fundamental image.
\end{enumerate}

Since typical coordinate grids are not mapped into themselves by the
fundamental group, the slave image points will not lie on the
computational mesh. Therefore we have to interpolate to find the field
value at each slave point. The procedure of point
sorting and image finding is illustrated in Fig.~2. Only one quadrant
of the FD is shown since the FD has 8-fold symmetry. The slave images
are those of slave points located in the other three quadrants.

\
\begin{figure}[h]\label{orbs}
\vspace*{60mm}
\includegraphics{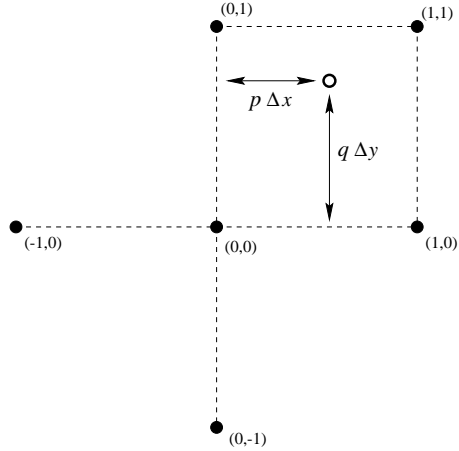}
\caption{Interpolating the image of a slave point.}
\end{figure}

The interpolation of the slave values is done using either a 3-point,
4-point or 6-point interpolation scheme. Employing the labelling
convention shown in Fig.~3, the 6-point interpolation is described by
\begin{eqnarray}\label{interp}
\phi(p\, \Delta x, q\, \Delta y) = &&q(q-1)/2\, \phi_{0 -1}
+ p(p-1)/2 \, \phi_{-1 0} \nonumber \\
&& +(1-pq-p^2-q^2)\phi_{00} + p(p-2q+1)/2\, \phi_{10} \nonumber \\
&& + q(q-2p+1)/2\, \phi_{0 1} +pq \phi_{1 1} + {\cal O}(\epsilon^3 \phi_{00})
\, .
\end{eqnarray}
Here $\epsilon={\rm max}(\Delta x, \Delta y)$.
By rotating the coordinates so that the interpolation is
performed in the quadrant $x>0,\, y>0$, we can ensure that the $(0,0)$
vertex of the cell enclosing the image point lies nearest the
origin. This increases the chances that $(-1,0)$ and $(0,-1)$ are
master points. Nevertheless, we need to check that all points involved
in the interpolation are either masters or slaves. If a freeman is
used in the interpolation it must be enslaved. Once this is done for
all the slave points we arrive at a set of $N$ coupled linear equations
involving $N$ slaves and $M$ masters (the slave drivers). Writing the
list of slaves and slave drivers as the column vectors $\phi_S$ and
$\phi_{SD}$, the interpolation equation (\ref{interp}) can be used to
form the linear system
\begin{equation}\label{slv}
A \phi_S = B \phi_{SD}\, \quad \Rightarrow \quad \phi_S = A^{-1}B
\phi_{SD} \, ,
\end{equation}
where $A$ is a $N \times N$ matrix and $B$ is a $N \times M$ matrix.
This procedure only has to be performed once at the beginning of a
simulation. The $N \times M$ matrix $C=A^{-1}B$ can then be stored and
called during the numerical evolution. In summary, at each time step
the masters are evolved according to (\ref{eveq}). The slaves are then
updated using (\ref{slv}). The slaves are the glue that holds together
identified sides of the fundamental cell. If the glue is poor energy
can either leak in or out of the system. Thus, by keeping track of the
total energy $E=K+G$ we can check that the periodic boundary
conditions are being properly implemented. The kinetic energy is given
by
\begin{equation}
K_k = \sum_{i,j}^{\rm masters} \left[ 
{ 4(\phi(i,j,k)-\phi(i,j,k-1) )^2 
\over (1-(i\Delta x)^2-(j \Delta y)^2)^2 \Delta t^2} \right]\Delta x
\Delta y \, .
\end{equation}
and the gradient energy is given by
\begin{equation}
\hspace*{-0.8in} G_k = \sum_{i,j}^{\rm masters}
\left[ { (\phi(i+1,j,k)-\phi(i,j,k) )^2 \over \Delta x^2}
+{ (\phi(i,j+1,k)-\phi(i,j,k) )^2 \over \Delta y^2} \right]\Delta
x\Delta y  ,
\end{equation}

\section{Results}

Several simulations were run using $301^2$ grids covering the region
$-0.8 < x,y < 0.8$. The coordinate grid spacings were: $\Delta x =
\Delta y = 2/375$ and $\Delta t = \Delta x /10$. The
proper distance between grid points varies between $\delta x = \Delta x$
near the origin and $\delta x = (2+\sqrt{2})\Delta x$ at the
extremities of the FD. Consequently, the resolution is $\sim 3.4$
times worse at the edge of the FD. Other metrics could be used to give
a more even coverage if desired. As a check on the accuracy of the
code the area was evaluated and compare to the continuum
answer:
\begin{equation}
A =\sum_{i,j}^{{\rm masters}} { 4 \Delta x \Delta y
\over (1-(i\Delta x)^2-(j \Delta y)^2)^2 } = (1.00014)\, 4\pi \, .
\end{equation}

Two types of initial conditions were used for $\phi(0,{\bf x})$. The
first employed a random superposition of Gaussian peaks and was
designed to excite as many eigenmodes as possible. The second employed
a superposition of two spherically symmetric eigenmodes
($q=0.1$ and $q=0.5$) of the Laplacian on $H^2$. 
This choice was designed to excite low lying eigenmodes.
The waveforms were smoothed to remove any large gradients caused by
the imposition of the periodic boundary conditions. The
residual monopole contribution was removed to increase our chances of
resolving any low lying multipoles. A stroboscopic sequence showing
the evolution of the random Gaussian peaks is shown in Fig.~4.

The wave was evolved for $2^{17}$ timesteps, which equates to
$T=69.9051$ curvature units, or roughly 20 light crossing times.
The total energy remained roughly constant throughout the simulation,
although there were $\sim \pm 2\%$ fluctuations about the mean value.
We believe these variations are due to the uneven grid resolution.

\newpage

\
\begin{figure}[h]
\vspace{120mm}

\includegraphics{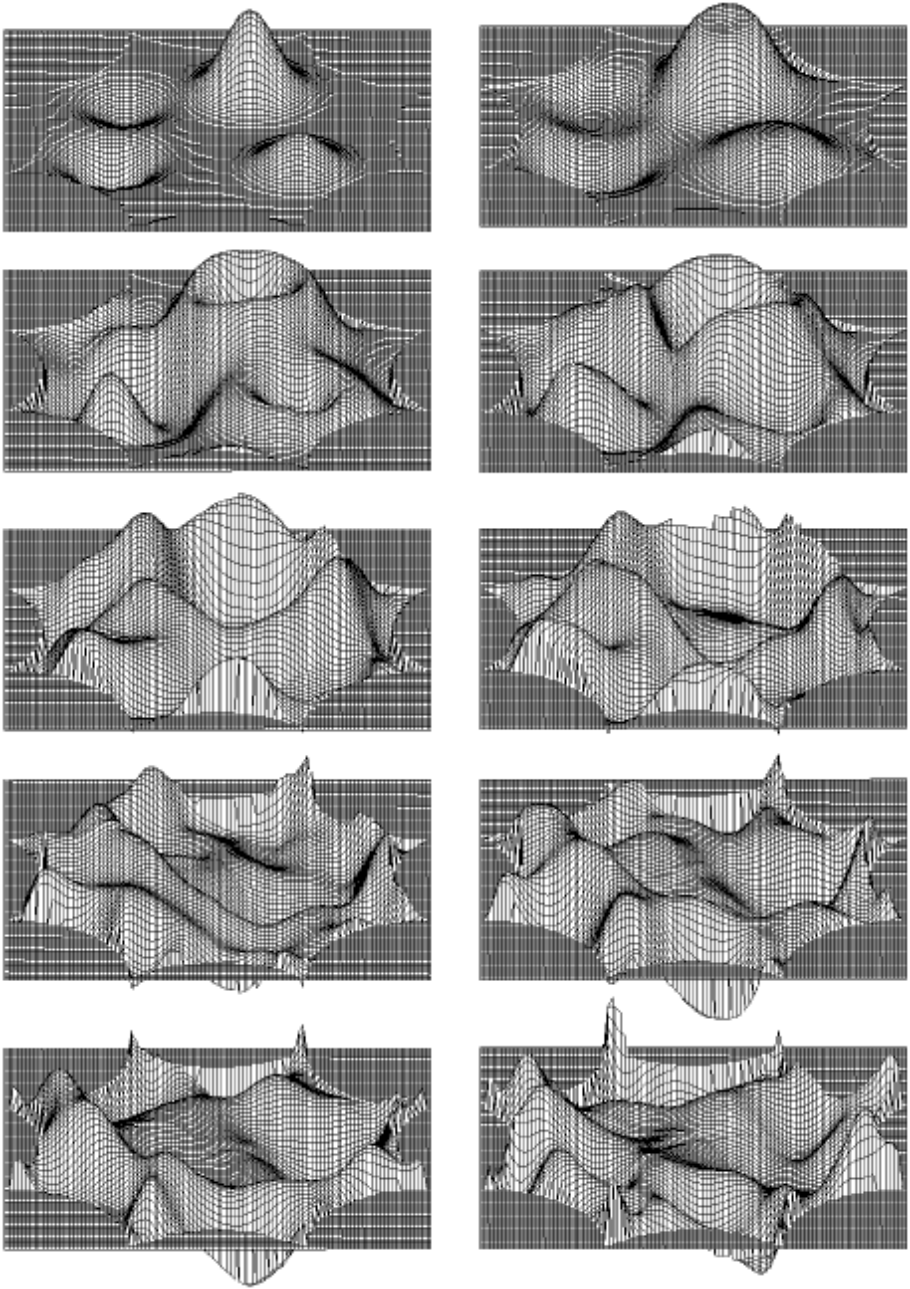}

\vspace{53mm}
\caption{The evolution of an initially static waveform shown
every 0.2 time units. Time runs cartoon style.}
\end{figure}

The wave was Fourier analysed to produce the scaled power spectrum shown in
Fig.~5. The power spectrum is multiplied by $\sqrt{\omega}$ for
$\omega>0$ to reflect the larger statistical weight of the high
frequency modes.

\
\begin{figure}[h]
\vspace*{58mm}
\includegraphics{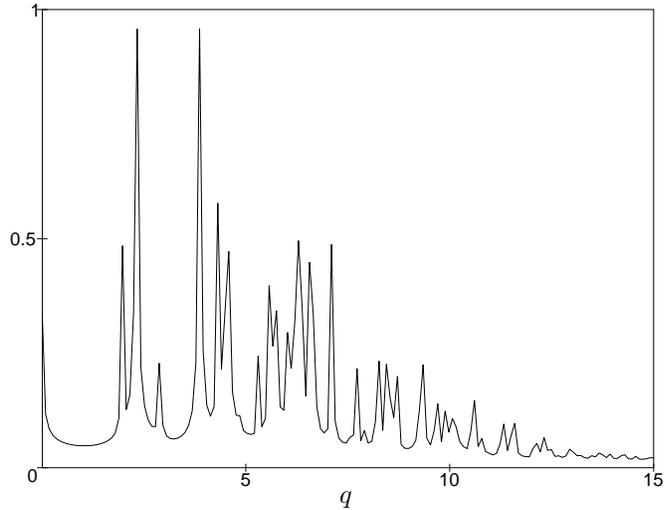}

\caption{The scaled power spectrum for the wave
shown in Fig.~4.}
\end{figure}

\vspace*{-2mm}
\begin{picture}(0,0)
\put(190,38){$q$}
\end{picture}

The first three peaks are at $\omega_1=q_1=1.97\pm 0.09$, $q_2=2.30\pm
0.09$ and $q_3=2.90\pm 0.09$. The errors reflect the finite frequency
resolution of $\Delta \omega = 2\pi/T = 0.09$. Aside from some
residual monopole power at $q=0$ there is no evidence for any low
lying eigenmodes. We will return to this crucially important point
later.

\
\begin{figure}[h]
\vspace*{55mm}
\includegraphics{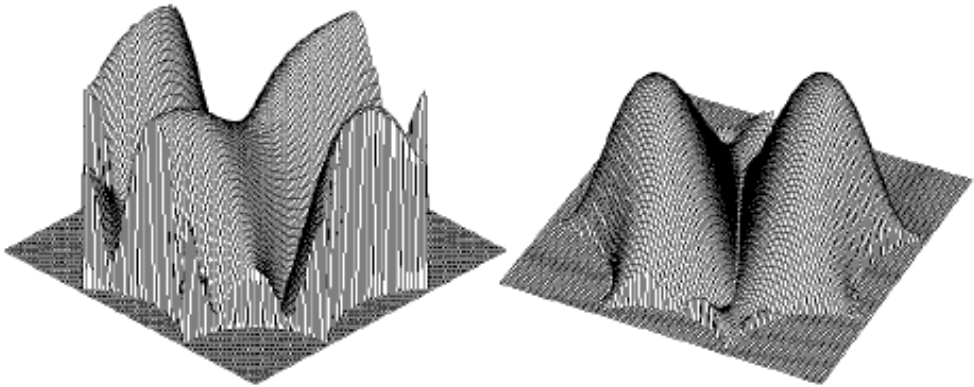}

\caption{The two lowest eigenmodes found by Fourier filtering.}
\end{figure}

By Fourier filtering at $\omega = \omega_1$ and $\omega= \omega_2$ we are able
to extract the corresponding spatial eigenmodes. These are displayed
in Fig.~6. By numerically evaluating the Laplacian of these
waveforms we were able to confirm that they are indeed eigenmodes
with the correct eigenvalues. We illustrate this in Fig.~7 by
displaying the eigenmode $\phi_{q_{3}}({\bf x})$ and
$-q_{3}^{-2} \nabla \phi_{q_{3}}({\bf x})$. The agreement is remarkably
good over most of the FD, save for some small regions where the spatial
gradients are large. The eigenmodes can be improved by a second
scrubbing run. To do this we use the approximate eigenmode as initial data
and evolve the system as before while filtering at the appropriate
eigenfrequency. This helps to remove contamination from the other
eigenmodes.

\newpage

\
\begin{figure}[h]
\vspace*{55mm}
\includegraphics{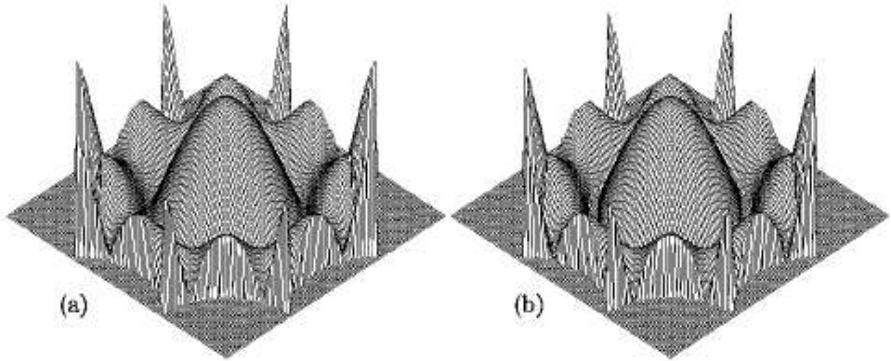}
\caption{(a) The $q_3$ eigenmode. (b) The $q_3$ eigenmode acted on
by $-q^{-2}_3\nabla$.}
\end{figure}

One possible explanation for the lack of eigenmodes below $q_1=1.97$
might be a lack of long range power in our choice of initial
conditions. To check this we tried a different initial waveform based
on a superposition of the spherically symmetric
$q=0.1$ and $q=0.5$ eigenmodes of $H^2$. When the periodic boundary
conditions are imposed the resulting initial waveform has 8-fold
symmetry.

\
\begin{figure}[h]
\vspace*{60mm}
\includegraphics{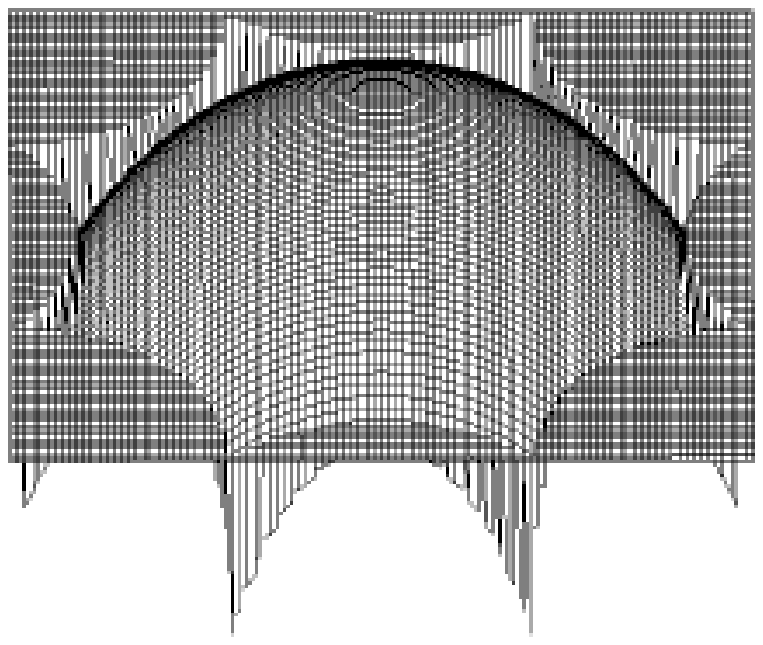}
\includegraphics{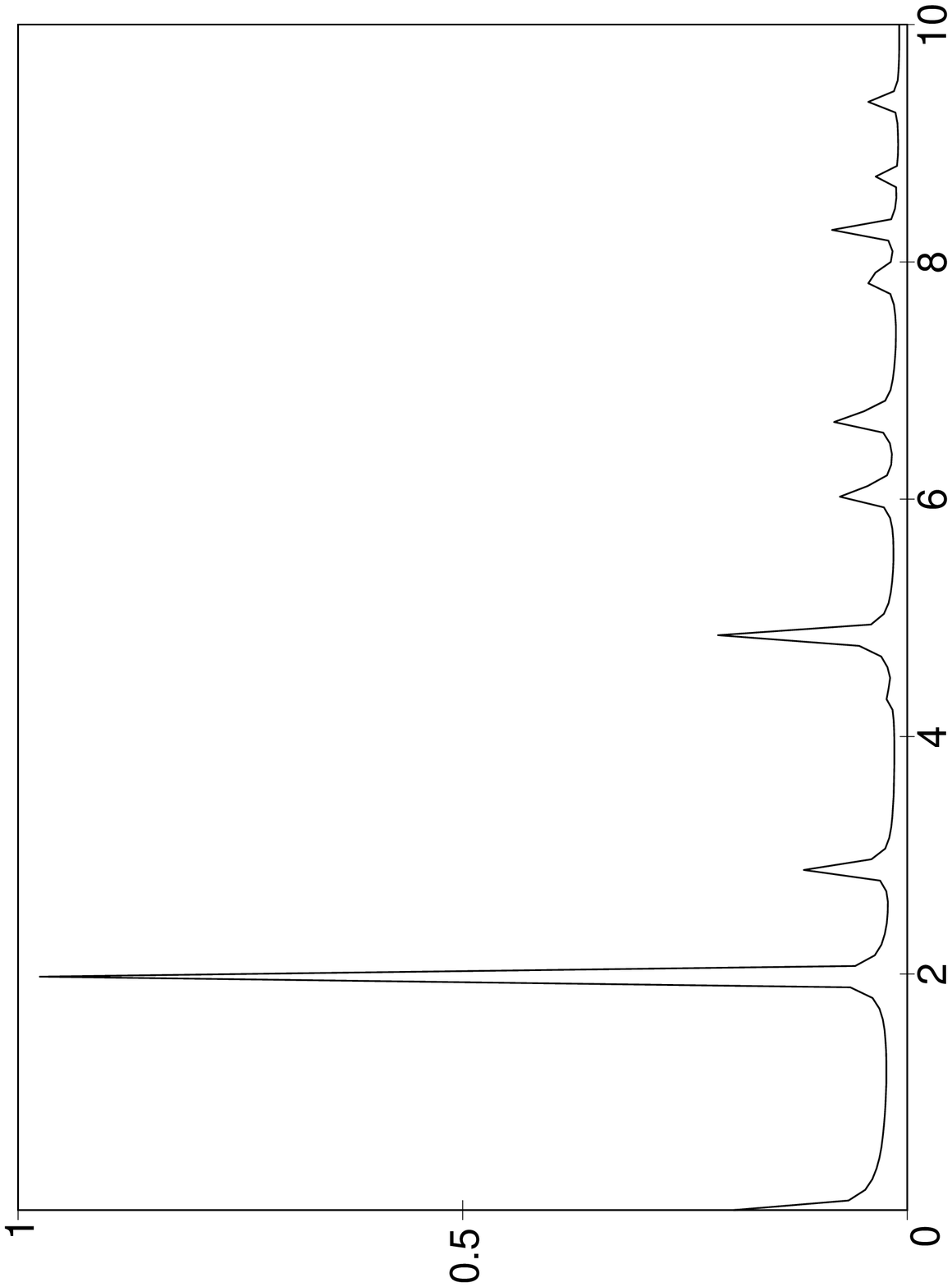}

\caption{The 8-fold symmetric $q=0.1,0.5$ initial waveform and the
resulting scaled power spectrum.}
\end{figure}

\vspace*{-2mm}
\begin{picture}(0,0)
\put(270,53){$q$}
\end{picture}

The scaled power spectrum shown in Fig.~8 shows no sign of any modes
below $q_1=1.97$. As expected, many of the modes excited by the random
superposition of Gaussian peaks are missing from the spectrum produced
by the 8-fold symmetric initial conditions. However, the modes that
are present agree with those in seen in Fig.~5.

\section{Supercurvature modes}

In $n$ dimensions the eigenmode spectrum for $H^n$ takes all values in
the range $q=[q_0,\infty)$, where $q_0=(n-1)/2$. Modes with $q<q_0$ are not
square integrable in infinite $n$-dimensional hyperbolic space, but
they are square integrable in the compact quotients
$\Sigma^n=H^n/\Gamma$. According to Buser\cite{buser}, many compact hyperbolic
manifolds support supercurvature modes with $q<q_0$. The name
supercurvature refers to the fact that these modes support significant
long range correlations on scales larger than the curvature scale.
In contrast, modes with $q>q_0$ have exponentially
damped correlations outside the curvature radius, even though their
wavelengths $\lambda = 2 \pi (q^2-q_0^2)^{-1/2}$ may greatly exceed the
curvature scale. This suppression can be understood on the grounds of
flux conservation in a space where the volume grows exponentially on
large scales. Supercurvature modes are very important in a
cosmological setting as even a single supercurvature mode can
significantly increase the amplitude of cosmic microwave background
fluctuations on large angular scales\cite{bell}.

There are a number of upper and lower bounds for the lowest
eigenvalue, $q_1>q_0$, of the Laplacian on $H^n/\Gamma$. In two
dimensions the tightest upper bound we know of was found by
Cheng\cite{cheng}:
\begin{equation}
q_1^2 \leq {1 \over 4} + \left({2 \pi \over d}\right)^2  \, ,
\end{equation}
where $d$ is the diameter of $\Sigma$. Applying this to our genus 2
example we find that $q_1\leq 2.61$, which is consistent with what we
found ($q_1=1.97\pm 0.09$). Using Cheeger's inequality, $q_1^2 \geq
h^2/4$, where $h$ is Cheeger's isoperimetric constant, Balazs \&
Voros\cite{balazs} suggest that $q_1 \geq 0.7$. This is also
consistent with our result.

To be certain that we have not missed any low lying eigenvalues we
need to use other methods such as the Selberg-Gutzwiller global
eigenvalue count\cite{gutz1}. After presenting our results in Cleveland, we
discovered a paper by Aurich \& Steiner\cite{as} that calculates all the
low lying eigenvalues for the double doughnut using a suitably
regularised Gutzwiller trace formula. They find the lowest three eigenvalues
to be $q_1=1.959$, $q_2=2.314$ and $q_3=2.872$, in perfect agreement
with our values. This proves that we did not miss any low lying
eigenvalues. While their values are sharper, the advantage of our
method is that it allows us to
extract the eigenmodes in addition to the eigenvalues.

The lack of low lying eigenmodes on the regular octagon is probably
related to its high degree of symmetry. Indeed, it is possible to move
to an alternative description of the manifold using a larger
fundamental group and a fundamental cell $96$ times smaller than
the octagon\cite{balazs}. Viewed from this perspective, the lowest
eigenmode is remarkably low, as it corresponds to a wave with
a wavelength $\sim 10$ times larger than the diameter
of the desymmetrised fundamental cell.

\section{Concluding Remarks}
Having demonstrated that our approach works in 2-dimensions,
our next task is to apply it to the cosmologically
relevant case where $\Sigma$ is a compact hyperbolic 3-manifold.
The computational cost of having an extra dimension will be offset by
the availability of examples with very small fundamental domains. Many small
hyperbolic 3-manifolds have FD's that have outradii,
$\eta_+$, smaller than the radius of curvature\cite{jeff}.
Consequently, space looks approximately Euclidean inside the FD, so
standard coordinate systems such as the Klein metric will produce
fairly uniform computational grids. Once we have found all
the low lying eigenmodes we can use them to produce simulated maps of
the cosmic microwave background radiation. These maps can then be
compared to observational data, or used to test proposals for finding
the large scale topology of the universe\cite{cssj}.

\section*{Acknowledgements}
We appreciate input from David Spergel, Glenn Starkman and Jeff Weeks.

\end{document}